\documentclass[aps,pre,twocolumn,superscriptaddress,showpacs]{revtex4}
\usepackage{graphicx,amsmath,
}
\begin{document}

\title{Spontaneous symmetry breaking of gap solitons in double-well traps}

\author{M.~Trippenbach}
\affiliation{Institute of Theoretical Physics, Physics Department, Warsaw University, Ho\.{z}a 69, PL-00-681
Warsaw, Poland}
\author{E. Infeld}
\affiliation{Soltan Institute for Nuclear Studies, Ho\.{z}a 69, PL-00-681 Warsaw, Poland}
\author{J. Goca{\l}ek}
\affiliation{Institute of Physics, Polish Academy of Sciences, Al. Lotników 32/46, Warsaw, Poland}
\author{Micha{\l} Matuszewski}
\affiliation{Nonlinear Physics Center and ARC Center of Excellence for
Quantum Atom Optics, Research School of Physical Sciences and
Engineering, Australian National University, Canberra ACT 0200,
Australia}
\author{M. Oberthaler}
\affiliation{Kirchhoff-Institut f\"ur Physik, Im Neuenheimer Feld 227, 69120 Heidelberg, Germany}
\author{B.~A.~Malomed.}
\affiliation{Department of Interdisciplinary Sciences, School of Electrical Engineering, Faculty of
Engineering, Tel Aviv University, Tel Aviv 69978, Israel}

\begin{abstract}
We introduce a two-dimensional model for the Bose-Einstein condensate with both attractive and repulsive nonlinearities. We assume a combination of a double-well potential in one direction, and an optical-lattice along the perpendicular coordinate. We look for dual-core solitons in this model, focusing on their symmetry-breaking bifurcations. The analysis employs a variational approximation, which is verified by numerical results. The bifurcation which transforms antisymmetric gap solitons into asymmetric ones is of supercritical type in the case of repulsion; in the attraction model, increase of the optical {\it latttice} strength leads to a gradual transition from subcritical bifurcation (for symmetric solitons) to a supercritical one.
\end{abstract}

\maketitle

\section{Introduction}

The Gross-Pitaevskii equation (GPE) provides a powerful model for studying the mean-field dynamics of Bose-Einstein condensates (BECs) \cite{GPE}. Important examples are the prediction of 1D gap solitons (GSs) in a self-repulsive condensate trapped in a periodic optical-lattice (OL) potential \cite{GSprediction}. This was realized experimentally in an ultracold gas of $^{87}$Rb atoms confined in a cigar-shaped trap \cite{GSexperiment}, and the prediction of the Josephson effect in a BEC \cite{JJT}.It was subsequently observed in a condensate trapped in a macroscopic double-well potential \cite{Markus}.  In contrast to hitherto realized Josephson systems in superconductors and superfluids, interactions between tunneling particles play a crucial role in a bosonic junction. The effective nonlinearity induced by the interactions gives rise to new effects in the tunneling. In particular, anharmonic Josephson oscillations were predicted \cite{JAVA,WALLS,SOLS}, provided that the initial population imbalance in the two potential wells falls below a critical value \cite{MIL,SMER}. This dynamic regime can be well explained by means of a simple model derived from the GPE, which amounts to a system of equations for the inter-well phase difference and population imbalance. The nonlinearity specific to the BEC also gives rise to a self-trapping effect in the form of a self-maintained population imbalance.

One-dimensional dynamics of a BEC in potentials composed of two rectangular potential wells were studied in several papers \cite{JJ1D}. Stationary states with different populations in the two wells are generated by symmetry-breaking bifurcations from symmetric and antisymmetric states, for attractive and repulsive nonlinearity, respectively \cite{MIL,SMER}. A natural 2D extension of the double-well configuration is a \textit{dual-channel} one, with the potential featuring the two wells in the direction of $x$, which are extended into parallel troughs along the $y$ axis \cite{first,Arik1}. In the case of an attractive nonlinearity, this setting may naturally give rise to dual-core solitons, which are self-trapped in the $y$ direction (similar to the ordinary matter-wave solitons created in a single-core trap \cite{solitons}), and are supported by a double-well structure in the perpendicular direction. Furthermore, if the nonlinearity is strong enough, or else the tunnel coupling between the troughs is weak, the obvious symmetric dual-core soliton may bifurcate into an \textit{asymmetric} one. This was demonstrated both in the full 2D model \cite{first}, and in its 1D counterpart, which replaces the 2D equation by a pair of one-dimensional GPEs with coordinate $y$, while the tunneling in the $x$ direction is approximated by a linear coupling between the equations \cite{Arik1}. In fact, the latter model resembles the standard one widely accepted in nonlinear optics to describe dual-core nonlinear optical fibers and asymmetric solitons \cite{dual-fiber,Progress}. In a similar way, the double-well potential may be uniformly extended in two transverse directions, giving rise to a 3D structure based on a pair of parallel ``pancakes".

If the dual-channel potential in 2D geometry is combined with an axial optical lattice, which runs along both potential troughs, it is natural to consider a dual-core gap soliton in the self-repulsive BEC filling this structure. In Ref. \cite{Arik1}, this was done using the above-mentioned approximation which replaced the corresponding two-dimensional GPE by a pair of linearly-coupled 1D equations. It was demonstrated that a symmetric gap solitons may be stable in this case, and never bifurcate, while asymmetric solitons are generated by a symmetry-breaking bifurcation from antisymmetric ones. Similar results (including the emergence of asymmetric gap solitons carrying intrinsic vorticity) where obtained in the 2D extension of the model. This model pertains to the above-mentioned ``dual-pancake" structure \cite{Arik2}. In nonlinear optics, asymmetric gap solitons were studied in models of dual-core fiber Bragg gratings, which also amount to systems of linearly coupled 1D equations \cite{MakTsofe}.

The prediction of symmetry breaking for matter-wave solitons in a setting combining the transverse double-well potential and a longitudinal optical lattice in experimentally relevant conditions makes it necessary to study the full 2D model (especially for the stability of the emerging asymmetric solitons) for both repulsive and attractive condensates, which is the purpose of the present work. Parameter regions admitting asymmetric solitons will be predicted by means of the variational approximation (VA) \cite{Progress}.  These results will be verified by numerics. The character of the symmetry-breaking bifurcations for the dual-core solitons will also be identified (we obtain a gradual transition from a subcritical bifurcation to a supercritical one with increase of the OL strength).

The paper is organized as follows. The model and the VA are introduced in Sec. II. In Sec. III we analyze the symmetry-breaking bifurcations in both attraction and repulsion models, and Sec. V concludes the paper.

\section{The model \ and variational approximation}

The normalized form of the GPE for the mean-field wave functions $\Psi $ in 2D geometry is
\begin{equation}
i\Psi _{t}=-(1/2)\left( \Psi _{xx}+\Psi _{yy}\right) +\left[ U(x)+\sigma
|\Psi |^{2}+\rho \cos \left( 2y\right) \right] \Psi ,  \label{eq1}
\end{equation}%
where $\sigma =+1$ and $-1$ for the self-repulsive and self-attractive BEC, and $\rho \cos \left( 2y\right) $ represents the longitudinal optical lattice potential. The transverse double-well structure is taken as%
\begin{equation}
U(x)=\left\{
\begin{array}{ll}
0, & |x|<L/2~\mathrm{and~}|x|~>L/2+D, \\
-U_{0}, & L/2<|x|<L/2+D,%
\end{array}%
\right.   \label{DC}
\end{equation}%
with $D$, $U_{0}$ and $L$ being, respectively, the width and depth of each well, and the width of the barrier between them, see Fig. \ref{potenlat} below.

Stationary solutions to Eq. (\ref{eq1}) are assumed in the form $\Psi
(x,y,t)=e^{-i\mu t}\Phi (x,y)$, where the real function $\Phi (x,y)$
satisfies the equation
\begin{equation}
\mu \Phi +(1/2)\left( \Phi _{xx}+\Phi _{yy}\right) -U(x)\Phi -\sigma \Phi
^{3}+\rho \cos (2y)\Phi =0.  \label{Phi}
\end{equation}%
It can be derived from the Lagrangian,
\begin{eqnarray}
L_{\mathrm{stat}} &=&\int \int dxdy\left[ \mu \Phi ^{2}-(1/2)\left( \Phi
_{x}^{2}+\Phi _{y}^{2}\right) -\right.   \notag \\
&&\left. -U(x)\Phi ^{2}-\left( \sigma /2\right) \Phi ^{4}+\rho \cos \left(
2y\right) \Phi ^{2}\right] .  \label{Lstat}
\end{eqnarray}%
To apply the VA, we follow Ref. \cite{first} and adopt an \textit{ansatz }%
consisting of two distinct parts. First, inside each potential trough, i.e., at $%
\left\vert x\mp \left( L+D\right) /2\right\vert <D/2$, the trial function is
\begin{equation}
\Phi _{\pm }(x,y)=A_{\pm }\cos \left( \pi \frac{x\mp \left( L+D\right) /2}{D}%
\right) \exp \left( -\frac{y^{2}}{2W^{2}}\right) ,  \label{inner}
\end{equation}%
where $A_{\pm }$ and $W$ are three variational parameters. This expression
implies different amplitudes and a common longitudinal width, $W$, of the
wave-function patterns in both troughs. In the $x$ direction, the ansatz (\ref%
{inner}) emulates the ground-state wave function in an infinitely deep
potential box, which vanishes at the edges of the trough, see Fig.~\ref{potenlat}.
In the $y$ direction, the ansatz approximates the self-trapped soliton by a Gaussian profile.
Outside the troughs (at $|x|>L/2+D$ and $|x|<L/2$), the ansatz also
follows the pattern of quantum mechanics, in the form of a superposition of
exponential wave functions:
\begin{equation}
\Phi (x,y)=\sum_{+,-}A_{\pm }\exp \left( -\sqrt{-2\mu }\left\vert x\mp \frac{%
L+D}{2}\right\vert -\frac{y^{2}}{2W^{2}}\right) ,  \label{outer}
\end{equation}%
with the same amplitudes $A_{\pm }$ and width $W$ as in Eq. (\ref{inner}).
The ansatz is not continuous at the edges of the troughs; however,
comparison with numerical findings (see Fig.~\ref{fig1} below) clearly
suggest that the VA can be used despite this local discrepancy.

\begin{figure}[tbp]
\includegraphics[width=7.5cm]{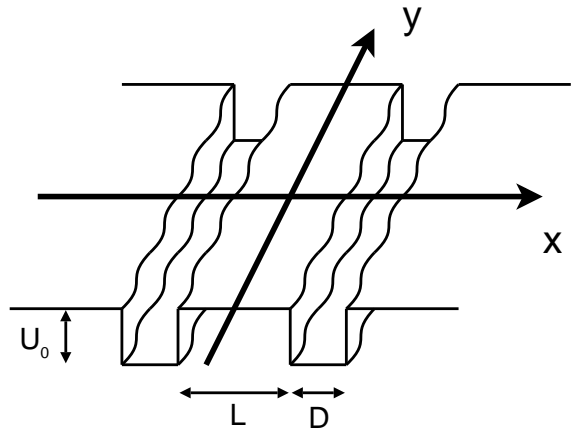}
\caption{(Color online) The shape of the quasi-one-dimensional double-well potential, $U(x,y)$. The wiggles indicate quasi-1D lattice along $y$.}
\label{potenlat}
\end{figure}

Substitution of expressions (\ref{inner}) and (\ref{outer}) into Eq. (%
\ref{Lstat}) and integration produce the following simplified Lagrangian, in which
contributions from the exponentially decaying functions in the outer region,
$|x|>L/2+D$, are neglected, the contribution from the optical lattice potential is taken
into account only inside the troughs, and the Thomas-Fermi approximation in
the $x$ direction is adopted, i.e., term $-(1/2)\Phi _{x}^{2}$ in the
Lagrangian density is omitted:
\begin{equation}
\frac{2}{D\sqrt{\pi }}L_{\mathrm{eff}}=\frac{1}{2}\rho We^{-W^{2}}\left(
A_{+}^{2}+A_{-}^{2}\right)
\end{equation}%
\begin{eqnarray}
+ &&\sum_{+,-}\left( \frac{\mu +U_{0}}{2}A_{\pm }^{2}W-\frac{A_{\pm }^{2}}{8W%
}-\frac{3\sigma }{2^{9/2}}A_{\pm }^{4}W\right)   \notag \\
&&+\frac{4\sqrt{-2\mu }}{D}e^{-\sqrt{-2\mu }\left( L+D\right) }A_{+}A_{-}W.
\label{L_eff}
\end{eqnarray}%
We now define $N_{\pm }\equiv \left( 3/4\sqrt{2}\right) A_{\pm }^{2}W$, and
\begin{equation}
\lambda \equiv \left( 2/D\right) \sqrt{-2\mu }\exp \left( -\sqrt{-2\mu }%
\left( L+D\right) \right) ,  \label{lambda}
\end{equation}%
\begin{equation}
N\equiv \frac{N_{+}+N_{-}}{4\sqrt{\lambda }},~\nu \equiv \frac{N_{+}-N_{-}}{4%
\sqrt{\lambda }},~\epsilon \equiv \mu +U_{0}.  \label{nuN}
\end{equation}%
The numbers of atoms trapped in the two troughs are proportional to the
respective partial norms of the wave function,
\begin{equation}
\left\vert \int_{-\infty }^{+\infty }dy\int_{\pm L/2}^{\pm \left(
D+L/2\right) }dx\left( \Phi (x,y)\right) ^{2}\right\vert =\frac{2\sqrt{2\pi }%
}{3}DN_{\pm }~,
\end{equation}%
hence $\nu $, defined in Eq. (\ref{nuN}), measures the \textit{population
imbalance. }In this notation, the Lagrangian (\ref{L_eff}) simplifies to
\begin{equation}
\frac{3}{8\sqrt{2\pi \lambda }D}L_{\mathrm{eff}}=
\end{equation}%
\begin{equation}
\equiv \frac{\epsilon N}{2}-\frac{N}{8W^{2}}-\sigma \frac{\sqrt{\lambda }}{2}%
\frac{N^{2}+\nu ^{2}}{W}-s\lambda \sqrt{N^{2}-v^{2}}+\frac{1}{2}\rho
Ne^{-W^{2}},  \label{Lfinal}
\end{equation}%
with $s=+1$ and $-1$ for the configurations of the antisymmetric and
symmetric types (with $A_{+}A_{-}<0$ and $A_{+}A_{-}>0$, respectively).

Our Lagrangian gives rise to variational equations $\partial
L/\partial W=\partial L/\partial \nu =\partial L/\partial N=0$:
\begin{eqnarray}
N+2\sigma \sqrt{\lambda }\left( N^{2}+\nu ^{2}\right) W-4\rho
NW^{4}e^{-W^{2}} &=&0,  \label{W} \\
\nu \left( -\frac{\sigma }{W}+s\sqrt{\frac{\lambda }{N^{2}-\nu ^{2}}}\right)
&=&0,  \label{nu} \\
\frac{1}{4W^{2}}+\sigma \frac{2\sqrt{\lambda }N}{W}+\frac{2s\lambda N}{\sqrt{%
N^{2}-\nu ^{2}}}-\rho e^{-W^{2}} &=&\epsilon .  \label{N}
\end{eqnarray}%
Equation (\ref{nu}) has two solutions: $\nu =0$, which corresponds to
symmetric or antisymmetric solitons, and
\begin{equation}
\nu ^{2}=N^{2}-\lambda W^{2},  \label{nu^2}
\end{equation}%
for asymmetric ones. Comparison of typical asymmetric and symmetric
solitons, found from a numerical solution of Eq. (\ref{Phi}), with their
counterparts predicted by the VA, is presented in Fig. \ref{fig1}.

\begin{figure}[tbp]
\includegraphics[scale=0.30]{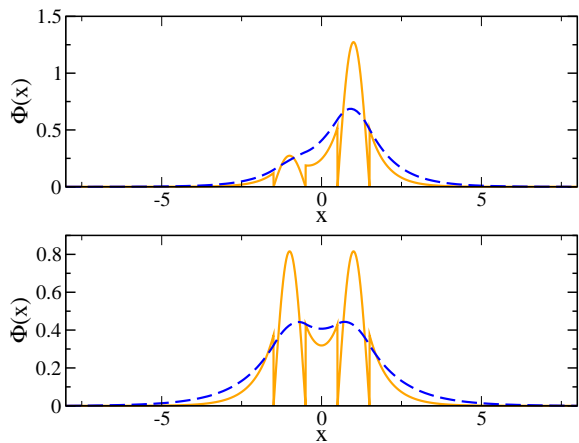}
\caption{(Color online) The top and bottom panels demonstrate examples of
cross-section profiles, along $y=0$, of stable asymmetric and symmetric gap
solitons in the model with repulsion, as obtained from a numerical solution
to Eq. (\protect\ref{Phi}) and predicted by the variational approximation
(dashed and continuous lines, respectively). Parameters of the double-well
potential are $L=D=1$, $U_{0}=-0.7$ (repulsive case) and $\rho=1$. Norms of the asymmetric and symmetric
solitons are, respectively, $N=0.52$ and $0.34$. The asymmetry parameter for
the former soliton, see Eqs. (\protect\ref{nuN}), is $\protect\nu =0.34$.}
\label{fig1}
\end{figure}

\begin{figure}[tbp]
\includegraphics[width=7.5cm]{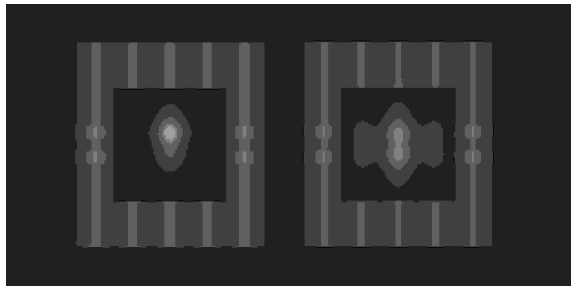}
\caption{Three dimensional version of the situation pictured in Fig.~\ref{fig1} . The density scale is represented  by the intensity of the print. Horizontal stipes represent double well structure and modulation illustrate the lattice} \label{fig_last}
\end{figure}

For symmetric and antisymmetric solitons, Eqs. (\ref{W}) and (\ref{N}), with
$\nu =0$, are tantamount to equations that were derived, by means
of the VA, for solitons in 1D models with a periodic sinusoidal potential
and attractive  or repulsive nonlinearity \cite{Wang,SKA}. In
particular, in the latter case (for $\sigma =+1$) a known fact is that
solutions exist only for $\rho >\rho ^{(0)}\equiv e^{2}/16\approx 0.462$ (in
fact, this constraint predicts, with high accuracy, the edge of the first
finite bandgap in the linear spectrum induced by the OL \cite{SKA}).
Results for asymmetric solitons are presented in the next section.

\section{Asymmetric solutions}

\subsection{Equations for the bifurcation point}

According to Eq. (\ref{nu}), asymmetric solutions exist in two cases: $%
\sigma =s=+1$ (repulsion, with the asymmetric branch bifurcating from the
antisymmetric one), or $\sigma =s=-1$ (attraction, with the bifurcation from
the symmetric branch). Elimination of $\nu ^{2}$ in Eqs. (\ref{W}) and (\ref%
{N}) by means of Eq. (\ref{nu^2}) yields a system of equations for $N$ and $W
$:%
\begin{eqnarray}
N+2\sigma \sqrt{\lambda }W\left( 2N^{2}-\lambda W^{2}\right)  &=&4\rho
NW^{4}e^{-W^{2}},  \notag \\
&&  \label{final} \\
\frac{1}{4W^{2}}+\sigma \frac{2\sqrt{\lambda }N}{W}+\frac{2s\sqrt{\lambda }N%
}{W}-\rho e^{-W^{2}} &=&\epsilon .  \notag
\end{eqnarray}%
Taking into account definitions (\ref{lambda}) and (\ref{nuN}), solutions to
Eqs. (\ref{final}) depend on parameters $L,D,U_{0}$, and $\rho $.

At the bifurcation point, $\nu =0$, Eq. (\ref{nu^2}) yields $N=\sqrt{\lambda
}W$, hence Eqs. (\ref{nu^2}) generate a system of two equations for two
coordinates of the bifurcation point, $\mu $ [via relations (\ref{nuN}) and (%
\ref{lambda})] and $W$:
\begin{eqnarray}
1+2\sigma \lambda W^{2} &=&4\rho W^{4}e^{-W^{2}},  \notag \\
&&  \label{bifpoint} \\
\frac{1}{4W^{2}}+2\left( \sigma +s\right) \lambda -\rho e^{-W^{2}}
&=&\epsilon .  \notag
\end{eqnarray}%
Without the OL, i.e., for $\rho =0$ (the case considered in Ref. \cite{first}%
), the first equation in (\ref{bifpoint}) gives the bifurcation point at $%
N=1/\sqrt{2}$. 
To obtain explicit results in the model with $\rho \neq 0$, one can start
with an obvious solution to Eqs. (\ref{bifpoint}), at $\lambda =\mu =N=0$, $%
\rho =\rho ^{(0)}$ (recall $\rho ^{(0)}\equiv e^{2}/16$), $%
U=U_{0}^{(0)}\equiv 1/16$, and $W=W^{(0)}\equiv \sqrt{2}$. This solution,
which has $N=0$ is, by itself, trivial, but a nontrivial one can be obtained
as an expansion around it.

\subsection{The model with self-attraction}

Consider the attraction model corresponding to $\sigma =s=-1$. Then,
straightforward analysis of Eqs. (\ref{bifpoint}) for small $\delta \rho
=\rho -\rho ^{(0)}$ and $\delta U_{0}=U_{0}-U_{0}^{(0)}$ demonstrates that
the bifurcation of symmetric solitons (which pertain to $s=-1$, see above)
may occur at two values of the norm,%
\begin{eqnarray}
N=\frac{1}{2\sqrt{2}}\sqrt{-\left( e^{-2}\delta \rho +\delta
U_{0}\right)}\nonumber \\
\left[ 2-\frac{1}{2}\left( e^{-2}\delta \rho +\delta U_{0}\right) \pm \sqrt{%
15e^{-2}\delta \rho -\delta U_{0}}\right] ,  \label{r12}
\end{eqnarray}%
the respective value of the width being $W\approx \sqrt{2}\left[ 1-\left(
e^{-2}\delta \rho +\delta U_{0}\right) /4\right] $. Note that the second
term in the square brackets in Eq. (\ref{r12}) is a small correction to $2$,
the main correction given by the last term, which demonstrates that theoretically there
may be two different bifurcation points. Obviously, expressions (\ref{r12})
are meaningful, i.e., the bifurcation takes place, if
\begin{equation}
\left( e^{2}/15\right) \delta U_{0}<\delta \rho <-e^{2}\delta U_{0}
\label{delta}
\end{equation}%
(in other words, $\delta U_{0}$ must be negative, while $\delta \rho $ may
have either sign). Numerical calculations imply that only the lower value of $N$ is valid.

\begin{figure}[tbp]
\includegraphics[width=6.5 cm]{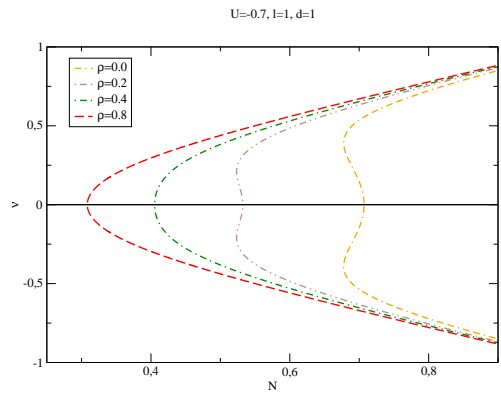}
\caption{(Color online)A set of numerically found bifurcation
diagrams in the model with attraction, showing degree of a
asymmetry of dual-core soliton, $\protect\nu $, as a function of
the soliton's total norm, $N$, see Eqs. (\protect\ref{nuN}). The
diagrams pertain to fixed values of parameters of the transverse
double-well configuration, $L=D=1$, $U_{0}=-0.7$ (attractive case), while the strength
of the longitudinal optical-lattice potential gradually increases. One can check from the analysis of Eq.~(\ref{bifpoint}) that the turning points are at $\nu =0$ and $\pm N/\sqrt{3}$. One can clearly see that the supercritical bifurcation will turn into subcritical bifurcation with increase of the optical {\it latttice} strength. }
\label{fig2}
\end{figure}

A set of bifurcation diagrams in the attraction model, in the form of $\nu
(N)$, i.e., curves showing the asymmetry of the dual-core solitons versus
the total norm, was generated by a numerical solution of the full system of
Eqs. (\ref{final}). The set is displayed in Fig. \ref{fig2}, where a
noteworthy feature is the transition from the \textit{subcritical} shape
(backward-directed one), which is a characteristic of the attraction
model without the longitudinal OL \cite{first} (as well as to the model of
dual-core optical fibers \cite{dual-fiber}), to the simpler \textit{%
supercritical} (forward-directed) shape at sufficiently large values of OL
strength $\rho $. Note that the symmetry-breaking bifurcations of dual-core
solitons, studied in systems of linearly-coupled GPEs including the
attractive nonlinearity and OL potential \cite{Arik1,Arik2}, as well as in
the system of linearly-coupled fiber Bragg gratings \cite{MakTsofe}, are of
supercritical type too. The physical significance of the subcritical
bifurcation is that it allows bistability of the solitons (the coexistence
of stable symmetric and asymmetric ones) in a limited interval of values
of $N$.

\subsection{The model with self-repulsive nonlinearity}

In the case of the self-repulsion, i.e., $\sigma =s=+1$, the expansion of
Eqs. (\ref{bifpoint}) predicts the following values of the norm at which
asymmetric gap solitons may bifurcate from the antisymmetric ones
(recall antisymmetric solitons corresponds to $s=+1$):
\begin{eqnarray}
N=\frac{1}{2\sqrt{2}}\sqrt{e^{-2}\delta \rho +\delta U_{0}}\nonumber\\
\left[ 2-\frac{1}{%
2}\left( e^{-2}\delta \rho +\delta U_{0}\right) \pm \sqrt{15e^{-2}\delta
\rho -\delta U_{0}}\right] ,  \label{r17}
\end{eqnarray}%
where the notation is the same as in Eq. (\ref{r12}) for the {\it attractive}
model. This expression predicts the bifurcation in the following region [cf.
Eq. (\ref{delta}) in the attraction model]: $-e^{-2}\delta \rho <\delta
U_{0}<15e^{-2}\delta \rho ,$ which implies $\delta \rho >0$, while $\delta
U_{0}$ may be both positive and negative, in contrast with the case of the
attraction model, that demanded $\delta U_{0}<0$, while allowing $\delta \rho $
to take either sign. Once again numerical calculations imply that only the lower value of $N$ is valid.

\begin{figure}[tbp]
\includegraphics[width=6.5 cm]{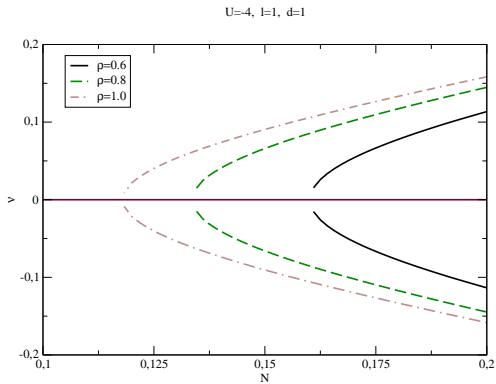}
\caption{(Color online) A set of bifurcation diagrams for gap
solitons in the model with repulsive nonlinearity, for $L=D=1$,
$U_{0}=4$, and a set of different values of the OL strength,
$\protect\rho $.} \label{fig3}
\end{figure}

A typical set of bifurcation diagrams in the repulsive model is displayed in
Fig. \ref{fig3}. It is seen that the bifurcation generating
asymmetric gap solitons from the antisymmetric ones is always of supercritical
type, in compliance with results obtained for the models based on linearly
coupled GPEs with the optical lattice potential and repulsive nonlinearity \cite%
{Arik1,Arik2}. These bifurcation diagrams exist only for $\rho >\rho
^{(0)}\equiv e^{2}/16$, because, as said above, at smaller values of the optical lattice
strength the VA does not predict antisymmetric GSs that might give rise to
a bifurcation.

\section{Conclusions}

We have introduced a 2D model for self-attractive and self-repulsive BECs, which combines a double-well potential in the transverse direction, and a periodic potential along the longitudinal coordinate. The  analysis involved symmetry-breaking bifurcations for dual-core solitons. Systematic results were obtained by means of the  variational approximation, which was verified by numerical results. In the case of a repulsive nonlinearity, the bifurcation is of supercritical type, while in the model with attraction an increase of the optical lattice strength leads to a gradual transition from subcritical bifurcation to a supercritical one. This is an important result.

\section{Acknowledgements}

M.T. acknowledges the support of the Polish Government Research Grant for 2006-2009. E.I and M.M. acknowledges the support of the Polish Government Research Grant for 2007-2010 and 2007+2009. The work of B.A.M.  was partially supported by the Israel Science Foundation through Excellence-Center grant No. 8006/03. He would like to thank Soltan Institute for Nuclear Studies, Warsaw, for an invitation in 2007. B.A.M. and M.O. acknowledge the support bz German-Israel Foundation through grant No. 149/2006.


\end{document}